\documentclass{colt}

\title{Accurate Community Detection in the Stochastic Block Model via Spectral Algorithms}
\usepackage{times}

\coltauthor{\Name{Se-Young Yun} \Email{seyoung.yun@inria.fr}\\
 \addr MSR-Inria, 23 Avenue d'Italie, 75013 Paris, France 
 \AND
 \Name{Alexandre Proutiere} \Email{alepro@kth.se}\\
 \addr KTH, The Royal Institute of Technology, EE School / ACL, Osquldasv. 10, Stockholm 100-44, Sweden
 }

\usepackage{amsmath,amssymb, amsfonts}
\usepackage{bm}
\usepackage{algorithm,algorithmic,bbm}

\newcommand{\Ex}{{\mathbb E}}

\begin{document} 

\maketitle

\begin{abstract}
We consider the problem of community detection in the Stochastic Block Model with a finite number $K$ of communities of sizes linearly growing with the network size $n$. This model consists in a random graph such that each pair of vertices is connected independently with probability $p$ within communities and $q$ across communities. One observes a realization of this random graph, and the objective is to reconstruct the communities from this observation. We show that under spectral algorithms, the number of misclassified vertices does not exceed $s$ with high probability as $n$ grows large, whenever $pn=\omega(1)$, $s=o(n)$ and 
\begin{equation*}
\lim\inf_{n\to\infty} {n(\alpha_1 p+\alpha_2 q-(\alpha_1 +
  \alpha_2)p^{\frac{\alpha_1}{\alpha_1 + \alpha_2}}q^{\frac{\alpha_2}{\alpha_1 + \alpha_2}})\over \log (\frac{n}{s})} >1,\quad\quad(1)
\end{equation*}
where $\alpha_1$ and $\alpha_2$ denote the (fixed) proportions of vertices in the two smallest communities. In view of recent work by \cite{abbe2014exact} and \cite{mossel2014consistency}, this establishes that the proposed spectral algorithms are able to exactly recover communities whenever this is at all possible in the case of networks with two communities with equal sizes. We conjecture that condition (1) is actually necessary  to obtain less than $s$ misclassified vertices asymptotically, which would establish the optimality of spectral method in more general scenarios.      
\end{abstract}

\section{Introduction}

Extracting structures or communities in networks is a central task in many disciplines including social sciences, biology, computer science, statistics, and physics. The Stochastic Block Model (SBM) was introduced a few decades ago as a performance benchmark to study the problem of community detection in random graphs, and it has, since then, attracted a lot attention. In this paper, we provide new results on the performance of spectral algorithms for detecting communities in the SBM. We consider a network consisting of a set $V$ of $n$ nodes. $V$ admits a hidden partition of $K$ non-overlapping subsets or communities $V_1,\ldots,V_K$ ($V=\bigcup_{k=1}^KV_k$). The size of community $V_k$ is $\alpha_k\times n$ for some $\alpha_k>0$. Without loss of generality, let $\alpha_1 \le \alpha_2 \le \dots \le \alpha_K$. We assume that when the network size $n$ grows large, the number of communities $K$ and their relative sizes are kept fixed. The communities have to be reconstructed from an observed realization of a random graph constructed as follows. Each pair of vertices is connected independently with probability $p$ within communities and $q$ across communities, where $p$ and $q$ may depend on the network size $n$. We assume that there exists $\epsilon>0$ such that ${p\over q}\ge 1+\epsilon$ uniformly in $n$. We further restrict our attention to the sparse case such that $p=o(1/\log^2 n)$ and the case where $pn=\omega(1)$, which is a necessary condition for asymptotically accurate community detection i.e., for the existence of algorithms that yield a vanishing proportion of misclassified vertices.

We show that under certain spectral algorithms, the number of misclassified vertices does not exceed $s$ with high probability as $n$ grows large, whenever $s=o(n)$ and 
\begin{equation}\label{eq:opt}
\lim\inf_{n\to\infty} {n(\alpha_1 p+\alpha_2 q-(\alpha_1 +
  \alpha_2)p^{\frac{\alpha_1}{\alpha_1 + \alpha_2}}q^{\frac{\alpha_2}{\alpha_1 + \alpha_2}})\over \log (\frac{n}{s})} >1.
\end{equation}

This result extends recent work about exact community reconstruction in the binary symmetric SBM (i.e., in the specific case of two communities of equal sizes). Indeed, by choosing $s<1$ in (\ref{eq:opt}), we get a condition under which spectral algorithms exactly recover the structure of any asymmetric networks with an arbitrary (but finite) number of communities. However our results is not limited to exact reconstruction, as we may choose any $s=o(n)$, e.g., $s=\sqrt{n}$. 

We conjecture that the condition (\ref{eq:opt}) is necessary for the existence of algorithms yielding less than $s$ misclassified vertices. The conjecture is true in the case of exact reconstruction ($s<1$) for the binary symmetric SBM. Please refer to the next section for a more detailed description on the related work.

\section{Previous Results}

{\bf Exact Detection.} Asymptotically exact community reconstruction in the SBM has been recently addressed in \cite{abbe2014exact}, \cite{mossel2014consistency}, and \cite{hajek2014achieving}. These papers only consider the binary symmetric SBM. They establish a necessary and sufficient condition for asymptotically exact reconstruction that coincides with (\ref{eq:opt}) when applied to two communities of equal sizes ($\alpha_1=\alpha_2=1/2$) and $s<1$. For example, when $p={a\log(n)\over n}$ and $q={b\log(n)\over n}$ for $a>b$, (\ref{eq:opt}) becomes equivalent to ${a+b\over 2}-\sqrt{ab}>1$. The three aforementioned papers further provide optimal algorithms, i.e., algorithms exactly recovering the network structure when this is possible. Note that in \cite{abbe2014exact}, and \cite{hajek2014achieving}, the proposed algorithms are based on SDP, and can be computationally expensive. In contrast, we prove that simple spectral algorithms are optimal. 

\medskip
\noindent
{\bf Asymptotically Accurate Detection.} Necessary and sufficient conditions for asymptotically accurate detection (i.e., the proportion of misclassified vertices vanishes when $n$ grows large) in the SBM has been derived in \cite{yun2014}. This condition is $n(p-q)^2/(p+q)=\omega(1)$. In the present paper, we provide results that fill the gap between exact detection and asymptotically accurate detection.    

\medskip
\noindent
{\bf Detectability.} In the sparse regime where $p,q=o(1)$, and for the binary symmetric SBM, the main focus recently has been on identifying the phase transition threshold (a condition on $p$ and $q$) for {\it detectability}: It was conjectured in \cite{decelle2011} that if $n(p-q)<\sqrt{2n(p+q)}$ (i.e., under the threshold), no algorithm can perform better than a simple random assignment of vertices to communities, and above the threshold, communities can partially be recovered. The conjecture was recently proved in \cite{mossel2012stochastic} (necessary condition), and \cite{massoulie2013} (sufficient condition). 

\medskip
\noindent
A more exhaustive list of papers related to the SBM can be found in \cite{yun2014}.

\section{Spectral Algorithms and Their Performance}

\begin{algorithm}[t!]
   \caption{Spectral Partition}
   \label{alg:partition}
\begin{algorithmic}
\STATE {\bfseries Input:} Observation matrix $A$.
\STATE  {\bf 1. Trimming.} Construct $A_{\Gamma}=(A_{vw})_{v,w\in \Gamma}$ where $\Gamma = \{v:
  \sum_{w \in V} A_{vw} \le 5K \frac{\sum_{(v,w) \in E } A_{vw}}{n} \}$.
\STATE {\bf 2. Spectral Decomposition.} Run Algorithm~\ref{alg:specg}  with input $A_{\Gamma}, \frac{ \sum_{(v,w) \in E } A_{vw} }{n^2}$, and output $(S_k)_{k=1,\ldots,K}$.
\STATE {\bf 3. Improvement.}
   \STATE $S^{(0)}_k \leftarrow S_k ,$ for all $k$
   \FOR{$i=1$ {\bfseries to} $\log n$ }
     \STATE $S^{(i)}_k \leftarrow \emptyset ,$ for all $k$
      \FOR{$v \in V$}
  \STATE Find $k^{\star} = \arg \max_{k} \{\sum_{w \in S^{(i-1)}_k} A_{vw} / |
  S^{(i-1)}_k | \} $ (tie broken uniformly at random)
   \STATE $S^{(i)}_{k^{\star}} \leftarrow S^{(i)}_{k^{\star}} \cup \{ v \}$
   \ENDFOR
   \ENDFOR
\STATE $\hat{V}_k \leftarrow S_k^{(i)}$, for all $k$
   \STATE {\bfseries Output:} $(\hat{V}_k)_{k=1,\ldots,K}$.
\end{algorithmic}
\end{algorithm}

\begin{algorithm}[t!]
   \caption{Spectral decomposition }
   \label{alg:specg}
\begin{algorithmic}
   \STATE {\bfseries Input:} $A_{\Gamma}, \frac{ \sum_{(v,w) \in E } A_{vw} }{n^2}$ 
\STATE $\hat{A} \leftarrow $ $K$-rank approximation of $A_{\Gamma}$
\FOR{$i=1$ {\bfseries to} $\log n$ }
\STATE $Q_{i,v} \leftarrow \{ w \in \Gamma :
\| \hat{A}_w  -\hat{A}_v\|^2 \le i \frac{\sum_{(v,w) \in E } A_{vw} }{100n^2} \} $
\STATE $T_{i,0}\leftarrow \emptyset$
\FOR{$k=1$ {\bfseries to} $K$ }
\STATE $v_k^{\star} \leftarrow \arg \max_{v} | Q_{i,v}\setminus \bigcup_{l=1}^{k-1} T_{i,l} |$ 
\STATE $T_{i,k} \leftarrow Q_{i,v_k^{\star}} \setminus \bigcup_{l=1}^{k-1} T_{i,l} $ and $\xi_{i,k} \leftarrow \sum_{v \in T_{i,k}}  \hat{A}_v/ |T_{i,k}| .$
\ENDFOR
\FOR{$v \in \Gamma \setminus ( \bigcup_{k=1}^K T_{i,k} )$}
\STATE $k^{\star} \leftarrow \arg \min_{k} \| \hat{A}_v -\xi_{i,k}\|$ and  $T_{i,k^{\star}} \leftarrow T_{i,k^{\star}} \cup \{v\}$
\ENDFOR
\STATE $r_i  \leftarrow \sum_{k=1}^K \sum_{v \in T_{i,k}} \| \hat{A}_v -\xi_{i,k}\|^2$
\ENDFOR

\STATE $i^{\star} \leftarrow \arg \min_{i} r_i.$
\STATE $S_k\leftarrow T_{i^\star,k}$ for all $k$
\STATE {\bfseries Output:} $(S_k)_{k=1,\ldots,K}$.
\end{algorithmic}
\end{algorithm}

The proposed algorithm, referred to as Spectral Partition, is the same as that in \cite{yun2014}, and is simple modifications of algorithms initially presented in \cite{coja2010}. In this paper, we present a more precise analysis of its performance than that of \cite{yun2014}. Let $A$ denote the observed random adjacency matrix. The algorithm consists in three steps. \\
{\bf 1. Trimming.} We first trim the adjacency matrix $A$, i.e., we keep the entries corresponding to a set $\Gamma$ of vertices whose degrees are not too large. More precisely, $\Gamma = \{v: \sum_{w \in V} A_{vw} \le 10 \frac{\sum_{(v,w) \in E } A_{vw}}{n} \} $. The resulting trimmed observation matrix is denoted by $A_\Gamma$. \\
{\bf 2. Spectral decomposition.} We then extract the communities from the spectral analysis of $A_\Gamma$.\\
{\bf 3. Improvement.} Finally, we further improve the estimated communities. After
the spectral decomposition step, the identified communities
$(S_k)_{k=1,2}$ are good approximations of the true communities. The
improvement is obtained by sequentially considering each vertex and by
moving it to the community with which it has the largest number of edges.\\
The pseudo-code of the algorithm is presented in Algorithms 1 and 2. The next theorem provides performance guarantees for the Spectral Partition algorithm.

\begin{theorem}\label{thm:algorithms} Assume that for $n$ large enough:
$$
n(\alpha_1 p+\alpha_2 q-(\alpha_1 +
  \alpha_2)p^{\frac{\alpha_1}{\alpha_1 + \alpha_2}}q^{\frac{\alpha_2}{\alpha_1 + \alpha_2}})-\frac{np}{\log np} \ge \log
  (\frac{n}{s}).
  $$
Then under the Spectral Partition algorithm, the number of misclassified vertices is less than $s$ with high probability.
\end{theorem}

By assumption, we have ${p\over q}\ge 1+\epsilon$, and $pn=\omega(1)$. We may deduce that:
$$
n(\alpha_1 p+\alpha_2 q-(\alpha_1+\alpha_2)p^{\frac{\alpha_1}{\alpha_1 + \alpha_2}}q^{\frac{\alpha_2}{\alpha_1 + \alpha_2}})=\omega(\frac{np}{\log np}).
$$
This can be proven using extensions of the weighted Arithmetic-Mean Geometric-Mean inequality.  From Theorem \ref{thm:algorithms}, we deduce that: if 
\begin{equation*}
\lim\inf_{n\to\infty} {n(\alpha_1 p+\alpha_2 q-(\alpha_1 +
  \alpha_2)p^{\frac{\alpha_1}{\alpha_1 + \alpha_2}}q^{\frac{\alpha_2}{\alpha_1 + \alpha_2}})\over \log (\frac{n}{s})} >1,
\end{equation*}
then the Spectral Partition algorithm yields less than $s$ misclassified vertices with high probability. We conclude this paper by exemplifying the condition (\ref{eq:opt}). Consider the binary symmetric SBM, with $p={a\log(n)\over n}$ and $q={b\log(n)\over n}$ for some $a>b$. 
\begin{itemize}
\item Exact reconstruction: with $s<1$, (\ref{eq:opt}) is equivalent to ${a+b\over 2}-\sqrt{ab}>1$, which also constitutes a necessary condition for exact reconstruction. Theorem \ref{thm:algorithms} then states that Spectral Partition is optimal for exact reconstruction, i.e., it extracts the communities exactly whenever this is at all possible.
\item Accurate reconstruction: choose $s=n^x$ for some $x\in (0,1)$. Then (\ref{eq:opt}) is equivalent to 
$$
{a+b\over 2}-\sqrt{ab}>1-x.
$$
\end{itemize}

\bibliography{reference}

\appendix

\section{Proof of Theorem~\ref{thm:algorithms}}

\subsection{Preliminaries}

In what follows, we use the standard matrix norm $\| A\|=\sup_{\|
  x\|_2=1}\| Ax\|_2$. We define $X_{\Gamma} = A_{\Gamma} - \mathbb{E}[A_{\Gamma}]$, where $A_\Gamma$ is the adjacency matrix obtained after trimming (Step 1 in Algorithm \ref{alg:partition}). We also denote by $e(v,S) = \sum_{w \in S} A_{vw}$ the total number of edges in the observed graph including node $v$ and a node from $S$. 
  
We first provide key intermediate results.   

\begin{lemma}[Lemma 8.5 of \cite{coja2010}]
With high probability, $\| X_{\Gamma} \| = O(\sqrt{np})$.
\label{lem:FOspectral}
\end{lemma}

The proof of Lemma \ref{lem:FOspectral} relies on arguments used in
the spectral analysis of random graphs, see \cite{feige2005spectral}. The next lemma provides a bound on the number of misclassified nodes after spectral decomposition applied to the trimmed matrix $A_\Gamma$, see Algorithm \ref{alg:specg}. 

\begin{lemma}[Lemma 15 of \cite{yun2014}] Assume that $|V \setminus \Gamma | = O( 1/p )$ and $\| X_{\Gamma} \| = O(\sqrt{np})$. Let $(S_k)_{1\le k\le K}$ denotes the output of Algorithm \ref{alg:specg}. With high probability, there exists a permutation $\sigma$ of $\{1,\ldots,K\}$ such that:
$$
|\bigcup_{k=1}^{K}(V_{\sigma(k)}\setminus S_k )\cap \Gamma|  = O(1/p).
$$
\label{lem:spect}
\end{lemma}

Observe that using Chernoff bound, we have $|V\setminus \Gamma| = O(1/p)$, hence combining the two previous lemma yields:

\begin{corollary} Assume that $np=\omega(1)$. 
The output $(S_k)_{1\le k\le K}$ of Algorithm \ref{alg:specg} satisfies: with high probability, there exists a permutation $\sigma$ of $\{1,\ldots,K\}$ such that $\frac{1}{n}|\bigcup_{k=1}^K V_k \setminus S_k| = O(\frac{1}{np}).$
\label{thm:specdecom}
\end{corollary}

\noindent{\bf Proof of Theorem~\ref{thm:algorithms}:}
Let $H$ be the largest set of vertices $v\in V$ satisfying:
\begin{itemize}
\item[(H1)] When $v\in V_k$, $\frac{e(v,V_k)}{|V_k|} -\frac{e(v,V_j)}{|V_j|} \ge
  \frac{p}{\log^4 np}$ for all $j \neq k$.
\item[(H2)] $e(v,V) \le 10 np$
\item[(H3)] $e(v,V\setminus H) \le 2 \log^2 np,$
\end{itemize}
The proof proceeds as follows. We first show that $| V\setminus H|\le s$ with high probability. To this aim, we control the number of vertices satisfying (H1), (H2), and (H3), see Lemma~\ref{lem:tight}, Claim 1 and Lemma \ref{lem:sizeH}, respectively. The result is summarised in Lemma \ref{lem:sizeH}. Next Lemma \ref{lem:improve} establishes that there is no misclassified vertices in $H$ with high probability, which concludes the proof.

\begin{lemma} For $v\in V_k$, and for all $j\neq k$,
$$\mathbb{P}\{ \frac{e(v,V_k)}{|V_k|} -\frac{e(v,V_j)}{|V_j|} \le \frac{p}{\log^4 np} \} \le
\exp(-n(\alpha_1p + \alpha_2q - (\alpha_1 + \alpha_2)
p^{\frac{\alpha_1}{\alpha_1 + \alpha_2}}q^{\frac{\alpha_2}{\alpha_1 + \alpha_2}})+\frac{np}{2\log np}).$$ \label{lem:tight}
\end{lemma}
From Lemma~\ref{lem:tight}, with high probability, the number of vertices that do not satisfy (H1) is less than $s/3$ when $n(\alpha_1p + \alpha_2q - (\alpha_1 + \alpha_2)
p^{\frac{\alpha_1}{\alpha_1 + \alpha_2}}q^{\frac{\alpha_2}{\alpha_1 +
    \alpha_2}}) - \log(n/s) - \frac{np}{2\log np} = \omega(1)$, since
\begin{multline*}\frac{\mathbb{E}\{\mbox{The number of vertices that
      do not satisfy (H1)}\}}{s/3} \cr
\le \frac{3n}{s} \exp(-n(\alpha_1p + \alpha_2q - (\alpha_1 + \alpha_2)
p^{\frac{\alpha_1}{\alpha_1 + \alpha_2}}q^{\frac{\alpha_2}{\alpha_1 +
    \alpha_2}})+\frac{np}{2\log np}) = o(1) .\end{multline*}

\medskip
\noindent
{\bf Claim 1.} From Chernoff bound, we can easily show that  $v$ does
not satisfy (H2) with probability at most $\exp(-5np)$ and thus, with
high probability, the number of vertices that do not satisfy (H2) is
less than $\frac{s}{10}$, since
$$ \frac{\mathbb{E}\{\mbox{The number of vertices that
      do not satisfy (H1)}\}}{s/10} \le \frac{10n}{s} \exp(-np) = o(1).$$

In Lemma~\ref{lem:sizeH}, we conclude that $V\setminus H \le s$ after showing the number of vertices that do not satisfy (H3) is less that $\frac{s}{2}$ with high probability.
\begin{lemma}
When $n(\alpha_1p + \alpha_2q - (\alpha_1 + \alpha_2)
p^{\frac{\alpha_1}{\alpha_1 + \alpha_2}}q^{\frac{\alpha_2}{\alpha_1 + \alpha_2}}) - \log(n/s) - \frac{np}{2\log np} = \omega(1)$,
$|V \setminus H| \le s$, with
high probability. 

\label{lem:sizeH}
\end{lemma}

Lemma~\ref{lem:improve} shows that when initial (after Algorithm 2) number of misclassified vertices is $O(1/p)$, $$\frac{\# \mbox{misclassified
      vertices in $H$ at $i+1$-th iteration}}{\# \mbox{misclassified
      vertices in $H$ at $i$-th iteration}} \le e^{-2}.$$
Since the initial number of misclassified vertices is negligible compared
to $n$ by Lemma~\ref{thm:specdecom}, after $\log n$ iterations, there is no misclassified vertice in $H$.

\begin{lemma} If ${| \bigcup_{k=1}^K (S^{(0)}_k \setminus V_k)\cap H| + |V \setminus
  H|} = O(1/p),$
$$ \frac{ | \bigcup_{k=1}^K (S^{(i+1)}_k \setminus V_k)\cap H|}{
  |\bigcup_{k=1}^K (S^{(i)}_k \setminus V_k)\cap H|} \le \frac{1}{\sqrt{np}}.$$ \label{lem:improve}
\end{lemma}







\subsection{Proof of Lemma~\ref{lem:tight}}
From Chernoff bound, we know that for all $1 \le t \le K$, 
\begin{equation}\mathbb{P}\{ e(v,V_t) \ge \alpha_t n p \log np  \} =
  o(\exp(-np)). \label{eq:1}\end{equation}
Using \eqref{eq:1},
\begin{align}
\mathbb{P}&\{\frac{e(v,V_k)}{|V_k|} - \frac{e(v,V_j)}{|V_j|} \le
  \frac{p}{\log^4 np} \} \cr 
=&\mathbb{P}\{ -p\log np \le \frac{e(v,V_k)}{|V_k|} - \frac{e(v,V_j)}{|V_j|} \le \frac{p}{\log^4
  np} \} + \mathbb{P}\{\frac{e(v,V_k)}{|V_k|} - \frac{e(v,V_j)}{|V_j|} < -p\log np \} \cr
\le& \mathbb{P}\{ -p\log np \le \frac{e(v,V_k)}{|V_k|} - \frac{e(v,V_j)}{|V_j|} \le \frac{p}{\log^4
  np} \}+ o(\exp(-np)) \cr
\le& np \log np \mathbb{P}\{ e(v,V_k) - \lfloor
  \frac{\alpha_k}{\alpha_j}e(v,V_j)\rfloor = \lfloor
  \frac{np}{\log^4 np} \rfloor \} + o(\exp(-np))\label{eq:ev1}\\
\le& \exp\left((\alpha_k + \alpha_j )n p^{\frac{\alpha_k}{\alpha_k +
  \alpha_j}} q^{\frac{\alpha_j}{\alpha_k+ \alpha_j}} -\alpha_k n p -
    \alpha_j n q +\frac{np}{2\log np} \right). \label{eq:ev2}
\end{align}
We conclude the proof by proving \eqref{eq:ev1} and \eqref{eq:ev2}.

\noindent{\bf Proof of \eqref{eq:ev1}:} Since $\mathbb{P}\{ e(v,V_k) -
\lfloor\frac{\alpha_k}{\alpha_j}e(v,V_j)\rfloor = x \} \le \mathbb{P}\{ e(v,V_k) - \lfloor
  \frac{\alpha_k}{\alpha_j}e(v,V_j)\rfloor = \lfloor
  \frac{np}{\log^4 np} \rfloor\}$ for $-\lceil \alpha_k np \log
  np\rceil  \le x \le \lceil\frac{\alpha_k
      np}{\log^4 np} \rceil$,
\begin{align*} 
\mathbb{P}\{ -p\log np \le \frac{e(v,V_k)}{|V_k|}& - \frac{e(v,V_j)}{|V_j|} \le \frac{p}{\log^2
  np} \} \cr
\le &\sum_{x=-\lceil \alpha_k np \log np\rceil }^{\lceil\frac{\alpha_k
      np}{\log^4 np} \rceil}\mathbb{P}\{ e(v,V_k) - \lfloor\frac{\alpha_k}{\alpha_j}e(v,V_j)\rfloor = x \}\cr
\le & np \log np \mathbb{P}\{ e(v,V_k) - \lfloor
  \frac{\alpha_k}{\alpha_j}e(v,V_j)\rfloor = \lfloor
  \frac{np}{\log^4 np} \rfloor\}.
\end{align*}

\noindent{\bf Proof of \eqref{eq:ev2}:}
 Let $x^{\star} = \lfloor  \frac{np}{\log^4 np} \rfloor.$ 
\begin{align}
  \mathbb{P}&\{ e(v,V_k) - \lfloor
              \frac{\alpha_k}{\alpha_j}e(v,V_j) \rfloor =
              x^{\star} \} \} - \mathbb{P}\{e(v,V_k) > 10 np   \}\cr
  &\le  \mathbb{P}\{e(v,V_k) -
    \lfloor \frac{\alpha_k}{\alpha_j}e(v,V_j) \rfloor = x^{\star},~ e(v,V_k) \le 10 np  \} \cr
  & \le \sum_{i=0}^{10 np}\sum_{\ell=0}^{\lfloor \alpha_j / \alpha_k \rfloor}{\alpha_k n \choose i +
    x^{\star}}{\alpha_j n \choose \lceil \frac{\alpha_j}{\alpha_k}i \rceil + \ell}
    \left(\frac{p}{1-p} \right)^{i+x^{\star}} \left(\frac{q}{1-q} \right)^{\lceil \frac{\alpha_j}{\alpha_k}i \rceil + \ell}
    (1-p)^{\alpha_k n}(1-q)^{\alpha_j n} \cr
  &\stackrel{(a)}{\le} \sum_{i=0}^{10 np}\sum_{\ell=0}^{\lfloor
    \alpha_j / \alpha_k \rfloor}\left(\frac{e\alpha_k
    np}{(i+x^{\star})(1-p)} \right)^{i+x^{\star}}
    \left(\frac{e\alpha_jnq}{(\lceil \frac{\alpha_j}{\alpha_k}i \rceil
    + \ell )(1-q)} \right)^{\lceil \frac{\alpha_j}{\alpha_k}i \rceil +
    \ell}\exp(-\alpha_k n p - \alpha_j n q) \cr
& \le \sum_{i=0}^{10 np}\sum_{\ell=0}^{\lfloor
    \alpha_j / \alpha_k \rfloor}\left(\frac{e\alpha_k
    np}{i+x^{\star}} \right)^{i}
    \left(\frac{e\alpha_jnq}{\lceil \frac{\alpha_j}{\alpha_k}i \rceil
    + \ell } \right)^{\frac{\alpha_j}{\alpha_k}i} \exp(\frac{np}{\log^2 np}) \exp(-\alpha_k n p - \alpha_j n q)\cr
& \le \sum_{i=0}^{10 np}\sum_{\ell=0}^{\lfloor
    \alpha_j / \alpha_k \rfloor}\left(\frac{e\alpha_k
    np}{i} \right)^{i}
    \left(\frac{e\alpha_jnq}{\frac{\alpha_j}{\alpha_k}i  }
  \right)^{\frac{\alpha_j}{\alpha_k}i} \exp(\frac{np}{\log^2
  np}) \exp(-\alpha_k n p - \alpha_j n q)\cr
& \le (10 np+1)(\lfloor
    \frac{\alpha_j }{ \alpha_k} \rfloor+1)\exp\left((\alpha_k +
  \alpha_j )n p^{\frac{\alpha_k}{\alpha_k +
  \alpha_j}} q^{\frac{\alpha_j}{\alpha_k+ \alpha_j}} \right)\exp(\frac{np}{\log^2  np}) \exp(-\alpha_k n p - \alpha_j n q)\cr
  &\le \exp\left((\alpha_k + \alpha_j )n p^{\frac{\alpha_k}{\alpha_k +
  \alpha_j}} q^{\frac{\alpha_j}{\alpha_k+ \alpha_j}} -\alpha_k n p -
    \alpha_j n q +\frac{np}{4\log np} \right), 
\label{eq:npq}
\end{align}
where $(a)$ stems from the inequality ${n \choose k} \le
(ne/k)^k$. Since $ \mathbb{P}\{e(v,V_k) > 10 np  \} \le o(\exp(-np))$ from
 Chernoff bound, \eqref{eq:npq} implies \eqref{eq:ev2}.

\subsection{Proof of Lemma~\ref{lem:sizeH}}
 Let $Z_1$ denote the set of vertices that do not satisfy at least one of (H1) and (H2). From Lemma~\ref{lem:tight} and Chernoff bound, $|Z_1| < \frac{s}{2} $ with high probability. 

Next we prove the following intermediate claim: there is no subset
$S\subset V$ such that $e(S,S) \ge s \log^2 np$ and $|S|= s$ with high probability. For any subset $S \in V$
such that $|S| = s,$ by Markov inequality,
\begin{eqnarray} 
\mathbb{P} \{ e(S,S) \ge s \log^2 np \} &\le&\inf_{t\ge  0} \frac{ \mathbb{E}[ \exp (e(S,S)t) ]  }{ st \log^2 np } \cr
&\le&  \inf_{t\ge 0} \frac{ \prod_{i=1}^{s^2/2} ( 1+ p \exp(t) )  }{ st \log^2 np } \cr
&\le & \inf_{t\ge 0} \exp \left( \frac{s^2p}{2} \exp (t) - st \log^2 np  \right) \cr
&\le &  \exp \left(   -   nps\big(\log np -
       \frac{s}{2n}\exp(\frac{np}{\log np})\big) \right)\cr 
&\le& \exp \left(   -   \frac{nps \log np}{2} \right),\label{eq:bndss}
\end{eqnarray}
where, in the last two inequalities, we have set $t= \frac{np}{\log np}$ and used the fact that: 
$\frac{n}{s} \ge \exp(\frac{np}{\log np}),$
which comes from the assumptions made in the theorem. 
Since the number of subsets $S \subset V$ with size $s$ is
${{n}\choose{s}} \le (\frac{e n}{s})^{s} ,$ from \eqref{eq:bndss}, we deduce:
\begin{align*} \mathbb{E}[| \{S : e(S,S) \ge s \log^2 np ~\mbox{and}~|S|= s \} |]  
&\le (\frac{e n}{s})^{s} \exp \left(   -   \frac{nps \log np}{2}
  \right) \cr
&= \exp \left( -s(\frac{np \log np}{2} - \log \frac{en}{s}) \right) \cr
&\le \exp\left(-\frac{nps \log np}{4}\right).\end{align*}
Therefore, by Markov inequality, we can conclude that there is no $S \subset V$ such that
$e(S,S) \ge s \log^2 np$ and $|S|= s$ with high probability.

To conclude the proof of the lemma, we build the following sequence of sets. Let $\{ Z(i) \subset V\}_{1\le i \le i^{\star}}$ be generated as follows:
\begin{itemize}
\item $Z(0)=Z_1$.
\item For $i \ge 1$, $Z(i) = Z(i-1) \cup \{v_i \}$ if there exists
  $v_i \in V$ such that
  $e(v_i , Z(i-1)) \ge 2 \log^2 np$ and
  $v_i \notin Z(i-1)$ and if there does not exist, the sequence ends.
 \end{itemize}
The sequence ends after the construction of $Z(i^\star)$.
By construction, every $v \in V\setminus Z(i^\star)$ satisfies the
conditions (H1), (H2), and (H3). Since $H$ is the largest set of
vertices satisfying (H1), (H2), and (H3), $|H| \ge |V\setminus Z(i^\star)|$.   

The proof is hence completed if we show that $|Z(i^\star)|< s$. Let
$t^{\star} =  s- |Z_1|$. If $i^{\star} \ge t^{\star},$
$|Z(t^{\star})| =  s$ and since $|Z_1 | \le \frac{s}{2}$,
$$e(Z(t^\star) , Z(t^\star))\ge \sum_{i=1}^{t^\star}e(v_i , Z(i-1))
\ge 2 t^\star \log^2 np \ge s \log^2 np,$$
However, from the previous claim, we know that with high probability,
all $S\subset V$ such that $|S| =  s$ have to satisfy $e(S,S) \le
s \log^2 np$. Therefore, with high
probability, $i^\star < t^\star$ and
$$|Z(i^\star )|=i^{\star}+|Z_1| < t^{\star} + |Z_1| =  s.$$

\subsection{Proof of Lemma~\ref{lem:improve}}

We use the notation: $\mu(v, S) = \Ex [e(v,S) ].$ Let
$\mathcal{E}_{jk}^{(i)} = (S^{(i)}_j \cap
V_k)\cap H$ and $\mathcal{E}^{(i)}= \bigcup_{j,k:j \neq k} \mathcal{E}_{jk}^{(i)}$.
At each improvement step, vertices move to a community with more connections to it. Thus,
$$\sum_{j,k:j\neq k}\sum_{v \in \mathcal{E}_{jk}^{(i+1)}} \frac{e(v,S^{(i)}_j)}{|S^{(i)}_j|} - \frac{e(v,S^{(i)}_k )}{|S^{(i)}_k|} \ge 0. $$
Since $|\mathcal{E}^{(i)}| = O(1/p)$ and $e(v,V) \le 10np$ when $v\in
H$,
\begin{align*}
0\le &\sum_{j,k:j\neq k}\sum_{v \in \mathcal{E}_{jk}^{(i+1)}} \frac{e(v,S^{(i)}_j)}{|S^{(i)}_j|} - \frac{e(v,S^{(i)}_k )}{|S^{(i)}_k|}      \le \sum_{j,k:j\neq k}\sum_{v \in \mathcal{E}_{jk}^{(i+1)}} \frac{e(v,S^{(i)}_j)}{|V_j|} - \frac{e(v,S^{(i)}_k )}{|V_k|}  + \frac{\log np}{n}|\mathcal{E}^{(i+1)}|.
\end{align*}
With the above inequality and (H1), we can bound
$\frac{|\mathcal{E}^{(i+1)}|}{|\mathcal{E}^{(i)}|}$ as follows:
\begin{align*}
&-\frac{\log np}{n}|\mathcal{E}^{(i+1)}| \le  \sum_{j,k:j\neq k}\sum_{v \in \mathcal{E}_{jk}^{(i+1)}} \frac{e(v,S^{(i)}_j)}{|V_j|} - \frac{e(v,S^{(i)}_k )}{|V_k|}\cr
\le & \sum_{j,k:j\neq k}\sum_{v \in \mathcal{E}_{jk}^{(i+1)}}
      \frac{e(v,V_j)}{|V_j|}
 - \frac{e(v,V_k )}{|V_k|} + \sum_{v \in \mathcal{E}^{(i+1)}} \frac{e(v, \mathcal{E}^{(i)}
      \cup H)}{\alpha_1 n} \cr
\stackrel{(a)}{\le} & -|\mathcal{E}^{(i+1)}|\frac{p}{\log^4 np} + \sum_{v \in
      \mathcal{E}^{(i+1)}} \frac{e(v, \mathcal{E}^{(i)})}{\alpha_1 n} + \sum_{v \in
      \mathcal{E}^{(i+1)}} \frac{e(v, H)}{\alpha_1 n} \cr
= & -|\mathcal{E}^{(i+1)}|\frac{p}{\log^4 np} + \sum_{v \in
      \mathcal{E}^{(i+1)}} \frac{\mu(v, \mathcal{E}^{(i)})}{\alpha_1 n}+\sum_{v \in
      \mathcal{E}^{(i+1)}} \frac{(e(v, \mathcal{E}^{(i)})-\mu(v,
    \mathcal{E}^{(i)}))}{\alpha_1 n} + \sum_{v \in
      \mathcal{E}^{(i+1)}} \frac{e(v, H)}{\alpha_1 n} \cr
\le & -|\mathcal{E}^{(i+1)}|\frac{p}{\log^4 np} +
      \frac{p|\mathcal{E}^{(i)}||\mathcal{E}^{(i+1)}|}{\alpha_1 n}+\sum_{v \in
      \mathcal{E}^{(i+1)}} \frac{(e(v, \mathcal{E}^{(i)})-\mu(v,
      \mathcal{E}^{(i)}))}{\alpha_1 n} + \sum_{v \in
      \mathcal{E}^{(i+1)}} \frac{e(v, H)}{\alpha_1 n}\cr
\stackrel{(b)}{\le} & -|\mathcal{E}^{(i+1)}|\frac{p}{\log^4 np} +
      \frac{p|\mathcal{E}^{(i)}||\mathcal{E}^{(i+1)}|}{\alpha_1 n}+\frac{\sqrt{|\mathcal{E}^{(i)}||\mathcal{E}^{(i+1)}|
      \|X_\Gamma \|}}{\alpha_1 n} + \sum_{v \in \mathcal{E}^{(i+1)}}
                      \frac{e(v, H)}{\alpha_1 n} \cr
\stackrel{(c)}{\le} & -|\mathcal{E}^{(i+1)}|\frac{p}{\log^4 np} +
      \frac{p|\mathcal{E}^{(i)}||\mathcal{E}^{(i+1)}|}{\alpha_1
                      n}+\frac{\sqrt{|\mathcal{E}^{(i)}||\mathcal{E}^{(i+1)}|np\log
                      np}}{\alpha_1 n} + \frac{2 |\mathcal{E}^{(i+1)}|
                      \log^2 np}{\alpha_1 n},
\end{align*}
where $(a)$ stems from (H1), $(b)$ stems from the fact that $\sum_{v \in
      \mathcal{E}^{(i+1)}} (e(v, \mathcal{E}^{(i)})-\mu(v,
    \mathcal{E}^{(i)})) = 1_{\mathcal{E}^{(i)}}^{T} \cdot X_{\Gamma}
    \cdot 1_{\mathcal{E}^{(i+1)}}$ where $1_S$ indicates the vector
    $v$-th value is 1 if $v\in S$ and 0 otherwise, and $(c)$ stems
    from (H3).
Since $|\mathcal{E}^{(i)}| = O(1/p)$, we conclude that 
$$\frac{|\mathcal{E}^{(i)}|}{|\mathcal{E}^{(i+1)}|} \le \frac{1}{\sqrt{np}}. $$

\end{document}